\documentclass[10pt,conference]{IEEEtran}
\IEEEoverridecommandlockouts

\usepackage{booktabs}
\usepackage{multirow}
\usepackage{graphicx}
\usepackage{subcaption}
\usepackage{xcolor}
\usepackage{listings}
\usepackage{balance}
\usepackage{enumitem}
\usepackage{tikz}
\usepackage{cite}
\usepackage{url}
\usepackage{hyperref}
\usetikzlibrary{arrows.meta, positioning, shapes.geometric, fit, backgrounds, calc}

\usepackage{fancyhdr}
\fancypagestyle{jpmc}{%
  \fancyhf{}%
  \fancyfoot[C]{\thepage}%
  \fancyfoot[R]{\footnotesize \textcopyright~2026 JP Morgan Chase \& Co.}%
}
\makeatletter
\let\ps@IEEEtitlepagestyle\ps@jpmc
\makeatother

\begin{document}

\title{From Translation to Superset: Benchmark-Driven Evolution of a Production AI Agent from Rust to Python}

\author{
  \IEEEauthorblockN{Jinhua Wang}
  \IEEEauthorblockA{LLM Suite Team\\JP Morgan Chase \& Co.\\jinhua.wang@jpmorgan.com}
  \and
  \IEEEauthorblockN{Biswa Sengupta}
  \IEEEauthorblockA{LLM Suite Team\\JP Morgan Chase \& Co.\\biswa.sengupta@jpmorgan.com}
}

\maketitle

\IEEEpubidadjcol

\renewcommand{\thefootnote}{}
\footnotetext{\textbf{Disclaimer:} This paper was prepared for informational purposes by the LLM Suite group of JP Morgan Chase and its affiliates (`JPMC') and is not a product of the Research Department of JP Morgan. JP Morgan makes no representation, warranty or undertaking whatsoever and disclaims all liability for the completeness, accuracy or reliability of the information contained herein. This document is not intended as investment research or investment advice, or a recommendation, offer or solicitation for the purchase or sale of any security, financial instrument, financial product or service, or to be used in any way for evaluating the merits of participating in any transaction, and shall not constitute a solicitation under any jurisdiction or to any person, if such solicitation under such jurisdiction or to such person would be unlawful.}
\footnotetext{\textcopyright~2026 JP Morgan Chase \& Co. All rights reserved.}
\renewcommand{\thefootnote}{\arabic{footnote}}

\begin{abstract}
Cross-language migration of large software systems is a persistent engineering challenge, particularly when the source codebase evolves rapidly. We present a methodology for \emph{LLM-assisted continuous code translation} in which a large language model translates a production Rust codebase (648K~LOC, 65 crates) into Python (41K~LOC, 28 modules), and public agent benchmarks serve as the objective function that drives iterative refinement. Our subject system is \textsc{Codex~CLI}, a production AI coding agent. We demonstrate that: (1)~the LLM-translated Python port resolves \textbf{59/80 SWE-bench Verified tasks (73.8\%)} versus the Rust original's \textbf{56/80 (70.0\%)}, and achieves \textbf{42.5\% accuracy on Terminal-Bench} versus Rust's 47.5\% (post-fix complete rerun), demonstrating near-parity on real-world agentic tasks with Python slightly ahead on SWE-bench and Rust slightly ahead on Terminal-Bench; (2)~benchmark-driven debugging---where failing tasks reveal API protocol mismatches, environment pollution, tool-availability gaps, a silent WebSocket empty-response failure mode present in both implementations, and a model-generated invalid content-item type that crashed the agent---is more effective than static testing alone at closing the parity gap; (3)~the translation architecture supports \emph{continuous upstream synchronisation}, enabling the Python port to absorb new Rust commits through an LLM-assisted diff-translate-test loop; and (4)~the Python port has evolved into a \emph{capability superset} of the Rust original through a \texttt{codex.enhancements} module that adds 30 feature-flagged extensions---multi-agent orchestration, semantic memory, persistent plans, cost tracking, IDE bridge, guardian safety assessment, voice mode, and more---absent from the Rust implementation, while a layered flag-resolution system preserves strict parity mode for head-to-head comparison. Our evaluation across code complexity, test coverage, runtime performance, and head-to-head agent benchmarking shows that for LLM-based agents where API latency dominates, Python's expressiveness yields a 15.9$\times$ code reduction with negligible performance cost, while the benchmark-as-objective-function methodology provides a principled framework for growing a cross-language port from parity into a first-class extended platform.
\end{abstract}

\section{Introduction}
\label{sec:intro}

Cross-language migration of large software systems is a recurring challenge in software engineering. Teams migrate codebases for many reasons---performance, ecosystem access, contributor accessibility---but the process is labour-intensive, error-prone, and typically a one-time effort. Once migration is complete, the source and target diverge permanently; upstream improvements in the original language must be manually re-ported or abandoned. This problem is especially acute for fast-moving projects where the upstream ships daily.

Recent advances in large language models (LLMs) suggest a different approach: \emph{continuous, LLM-assisted code translation} where an LLM performs the bulk of cross-language translation, and automated benchmarks serve as the objective function that validates correctness and drives iterative refinement. Rather than a one-shot migration, this methodology establishes a \emph{living bridge} between the source and target codebases.

In this paper, we apply this methodology to \textsc{Codex CLI}~\cite{chen2021codex}, a production AI coding agent originally implemented in Rust (648K LOC across 65 crates). We translate it to Python using LLM-assisted translation, producing a 41K LOC implementation across 28 modules---a 15.9$\times$ code reduction. Crucially, we validate the translation not only through 2,621 unit tests but through \emph{head-to-head agent benchmarking}: running both the original Rust CLI and our Python port on Terminal-Bench~\cite{terminalbench2025}, an 80-task benchmark of complex terminal operations.

Our contributions are:

\begin{itemize}[leftmargin=*, nosep]
    \item \textbf{Benchmark-as-objective-function methodology.} We demonstrate that public agent benchmarks (Terminal-Bench, SWE-bench) serve as effective objective functions for cross-language translation. Our Python port achieves 42.5\% on Terminal-Bench versus the original's 47.5\%, and 73.8\% on SWE-bench Verified versus Rust's 70.0\%---near-parity on both benchmarks---with benchmark-driven debugging revealing protocol mismatches, environment pollution, tool-availability gaps, and transport-layer failure modes that unit tests missed.

    \item \textbf{From parity to superset.} We show that the same LLM-assisted translation methodology that achieves parity can continue to evolve the port beyond the original. The Python port ships a \texttt{codex.enhancements} module with 30 feature-flagged extensions---multi-agent orchestration, semantic memory, guardian safety assessment, cost tracking, and more---absent from the Rust implementation, while preserving a strict parity mode for fair comparison.

    \item \textbf{Continuous upstream synchronisation architecture.} We describe an LLM-assisted diff-translate-test loop that enables the Python port to continuously absorb upstream Rust commits. The architecture uses git submodule tracking, automated diff extraction, LLM-driven translation of changed modules, and benchmark regression testing to maintain parity as the upstream evolves.

    \item \textbf{Comprehensive empirical evaluation.} We evaluate the migration across nine dimensions: code size, cyclomatic complexity, test parity, runtime performance, API surface, dependency structure, migration effort, end-to-end agent benchmarks, and head-to-head benchmarking on both Terminal-Bench and SWE-bench Verified.

    \item \textbf{LLM-assisted translation patterns.} We document the systematic patterns by which an LLM translates Rust idioms to Python equivalents, including error handling (\texttt{Result<T,E>} $\to$ exceptions), concurrency (Tokio $\to$ asyncio), and serialization (serde $\to$ Pydantic), achieving idiomatic target code rather than mechanical transpilation.
\end{itemize}

Our findings have broad implications beyond this specific migration. The benchmark-as-objective-function approach is language-agnostic and applicable to any translation where functional equivalence can be measured through automated evaluation. For the emerging class of API-latency-bound LLM applications, our results demonstrate that Python's expressiveness advantages substantially outweigh Rust's performance benefits, yielding a more maintainable and extensible system.

The remainder of this paper is organized as follows: Section~\ref{sec:background} provides background on \textsc{Codex CLI} and AI coding agents. Section~\ref{sec:architecture} details the system architecture. Section~\ref{sec:migration} describes the LLM-assisted translation methodology and continuous sync architecture. Section~\ref{sec:evaluation} presents our empirical evaluation including head-to-head benchmark results. Section~\ref{sec:related} surveys related work, and Section~\ref{sec:conclusion} concludes.

\section{Background and System Overview}
\label{sec:background}

\subsection{AI Coding Agents}

AI coding agents are LLM-powered systems that interact with a developer's local environment to accomplish software engineering tasks. Unlike code completion tools that generate snippets, coding agents operate in a \emph{loop}: they analyze the current state of a codebase, formulate a plan, invoke tools (shell commands, file edits, web searches), observe results, and iterate until the task is complete or a human intervenes.

\textsc{Codex CLI} is a production AI coding agent that runs as a terminal application. It connects to OpenAI's Responses API, supports multiple LLM backends (GPT-4, o3, o4-mini, and local models via Ollama/LM Studio), and provides both interactive chat and non-interactive batch execution modes. The system enforces configurable security policies through platform-native sandboxing and supports extensibility through the Model Context Protocol (MCP)~\cite{zheng2025mcp}.

\subsection{System Scope}

The system comprises 28 Python modules (Table~\ref{tab:loc}), organized into six architectural layers:

\begin{enumerate}[leftmargin=*, nosep]
    \item \textbf{Agent Layer}: Core agent runner, tool orchestration, guardian (automated approval), multi-agent coordination, and memory management.
    \item \textbf{Security Layer}: Sandbox enforcement (Seatbelt, Bubblewrap, seccomp), execution policy evaluation, and process hardening.
    \item \textbf{Protocol Layer}: Wire protocol types, approval workflows, model abstractions, and JSON-RPC event system.
    \item \textbf{Integration Layer}: MCP client/server, authentication (OAuth/PKCE), backend API client, and cloud task management.
    \item \textbf{Presentation Layer}: Terminal UI (Textual framework), application server (WebSocket/HTTP), and CLI entry points.
    \item \textbf{Infrastructure Layer}: State persistence (SQLite), configuration management, telemetry, analytics, feature flags, and utilities.
\end{enumerate}

\subsection{Original Rust Implementation}

The system was originally implemented in Rust using the Tokio async runtime, comprising 65 crates with approximately 648K lines of code. The Rust implementation leveraged the language's type system for compile-time safety guarantees, particularly in the sandbox and protocol subsystems. However, the build system complexity (Bazel + Cargo), compilation times, and the cognitive overhead of Rust's ownership model motivated an exploration of alternative implementation languages.

\subsection{Migration Context}

The migration to Python was motivated by three factors: (1)~\emph{iteration velocity}---the ability to rapidly prototype and deploy agent capabilities is critical in a fast-evolving LLM landscape; (2)~\emph{ecosystem integration}---Python's dominance in the ML/AI ecosystem simplifies integration with model providers, data processing tools, and community extensions; and (3)~\emph{contributor accessibility}---Python lowers the barrier for external contributors and plugin developers.

\section{System Architecture}
\label{sec:architecture}

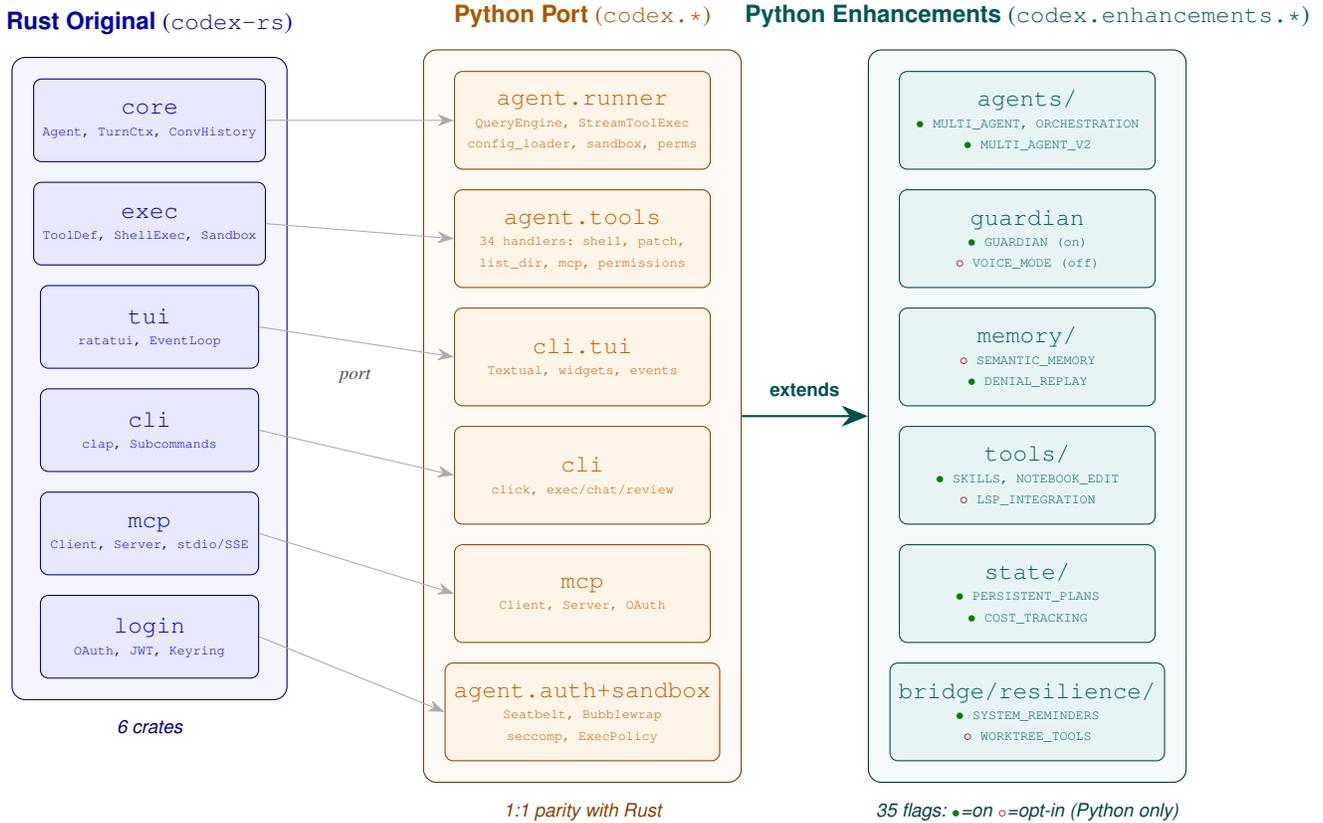
\begin{figure*}[t]
\centering
\begin{tikzpicture}[
    node distance=0.26cm,
    rcbox/.style={draw=blue!55!black, fill=blue!9, rounded corners=3pt,
        minimum width=2.9cm, minimum height=1.1cm,
        font=\small\ttfamily, text=blue!75!black, align=center},
    pybox/.style={draw=orange!55!black, fill=orange!9, rounded corners=3pt,
        minimum width=3.4cm, minimum height=1.3cm,
        font=\small\ttfamily, text=orange!75!black, align=center},
    enbox/.style={draw=teal!55!black, fill=teal!9, rounded corners=3pt,
        minimum width=3.4cm, minimum height=1.3cm,
        font=\small\ttfamily, text=teal!75!black, align=center},
    hdr/.style={font=\small\bfseries\sffamily},
    sub/.style={font=\scriptsize\itshape\sffamily},
    maparr/.style={-{Stealth[length=2.2mm,width=1.6mm]}, gray!65, thin},
    extarr/.style={-{Stealth[length=3.5mm,width=2.8mm]}, teal!60!black, thick},
]

\node[rcbox]                    (r1) {\texttt{core}\\[-2pt]{\tiny Agent, TurnCtx, ConvHistory}};
\node[rcbox, below=of r1]       (r2) {\texttt{exec}\\[-2pt]{\tiny ToolDef, ShellExec, Sandbox}};
\node[rcbox, below=of r2]       (r3) {\texttt{tui}\\[-2pt]{\tiny ratatui, EventLoop}};
\node[rcbox, below=of r3]       (r4) {\texttt{cli}\\[-2pt]{\tiny clap, Subcommands}};
\node[rcbox, below=of r4]       (r5) {\texttt{mcp}\\[-2pt]{\tiny Client, Server, stdio/SSE}};
\node[rcbox, below=of r5]       (r6) {\texttt{login}\\[-2pt]{\tiny OAuth, JWT, Keyring}};

\begin{scope}[on background layer]
  \node[draw=blue!40!black, fill=blue!4, rounded corners=5pt, inner sep=0.28cm,
        fit=(r1)(r2)(r3)(r4)(r5)(r6)] (Rpanel) {};
\end{scope}
\node[hdr, text=blue!68!black, above=5pt of Rpanel.north, anchor=south]
    {\textbf{Rust Original}~\normalfont\small(\texttt{codex-rs})};
\node[sub, text=blue!52!black, below=4pt of Rpanel.south, anchor=north]
    {6 crates};

\node[pybox, right=2.5cm of r1] (p1) {\texttt{agent.runner}\\[-2pt]{\tiny QueryEngine, StreamToolExec}\\[-2pt]{\tiny config\_loader, sandbox, perms}};
\node[pybox, below=of p1]       (p2) {\texttt{agent.tools}\\[-2pt]{\tiny 34 handlers: shell, patch,}\\[-2pt]{\tiny list\_dir, mcp, permissions}};
\node[pybox, below=of p2]       (p3) {\texttt{cli.tui}\\[-2pt]{\tiny Textual, widgets, events}};
\node[pybox, below=of p3]       (p4) {\texttt{cli}\\[-2pt]{\tiny click, exec/chat/review}};
\node[pybox, below=of p4]       (p5) {\texttt{mcp}\\[-2pt]{\tiny Client, Server, OAuth}};
\node[pybox, below=of p5]       (p6) {\texttt{agent.auth+sandbox}\\[-2pt]{\tiny Seatbelt, Bubblewrap}\\[-2pt]{\tiny seccomp, ExecPolicy}};

\begin{scope}[on background layer]
  \node[draw=orange!40!black, fill=orange!4, rounded corners=5pt, inner sep=0.28cm,
        fit=(p1)(p2)(p3)(p4)(p5)(p6)] (Ppanel) {};
\end{scope}
\node[hdr, text=orange!68!black, above=5pt of Ppanel.north, anchor=south]
    {\textbf{Python Port}~\normalfont\small(\texttt{codex.*})};
\node[sub, text=orange!52!black, below=4pt of Ppanel.south, anchor=north]
    {1:1 parity with Rust};

\foreach \s/\t in {r1/p1, r2/p2, r3/p3, r4/p4, r5/p5, r6/p6}{%
  \draw[maparr] (\s.east) -- (\t.west);
}
\node[font=\scriptsize\itshape, text=gray!60!black]
    at ($(Rpanel.east)!0.5!(Ppanel.west)$) [yshift=8pt] {port};

\node[enbox, right=2.5cm of p1] (e1) {\texttt{agents/}\\[-2pt]{\tiny \textcolor{green!50!black}{$\bullet$} MULTI\_AGENT, ORCHESTRATION}\\[-2pt]{\tiny \textcolor{green!50!black}{$\bullet$} MULTI\_AGENT\_V2}};
\node[enbox, below=of e1]       (e2) {\texttt{guardian}\\[-2pt]{\tiny \textcolor{green!50!black}{$\bullet$} GUARDIAN (on)}\\[-2pt]{\tiny \textcolor{red!60!black}{$\circ$} VOICE\_MODE (off)}};
\node[enbox, below=of e2]       (e3) {\texttt{memory/}\\[-2pt]{\tiny \textcolor{red!60!black}{$\circ$} SEMANTIC\_MEMORY}\\[-2pt]{\tiny \textcolor{green!50!black}{$\bullet$} DENIAL\_REPLAY}};
\node[enbox, below=of e3]       (e4) {\texttt{tools/}\\[-2pt]{\tiny \textcolor{green!50!black}{$\bullet$} SKILLS, NOTEBOOK\_EDIT}\\[-2pt]{\tiny \textcolor{red!60!black}{$\circ$} LSP\_INTEGRATION}};
\node[enbox, below=of e4]       (e5) {\texttt{state/}\\[-2pt]{\tiny \textcolor{green!50!black}{$\bullet$} PERSISTENT\_PLANS}\\[-2pt]{\tiny \textcolor{green!50!black}{$\bullet$} COST\_TRACKING}};
\node[enbox, below=of e5]       (e6) {\texttt{bridge/resilience/}\\[-2pt]{\tiny \textcolor{green!50!black}{$\bullet$} SYSTEM\_REMINDERS}\\[-2pt]{\tiny \textcolor{red!60!black}{$\circ$} WORKTREE\_TOOLS}};

\begin{scope}[on background layer]
  \node[draw=teal!40!black, fill=teal!4, rounded corners=5pt, inner sep=0.28cm,
        fit=(e1)(e2)(e3)(e4)(e5)(e6)] (Epanel) {};
\end{scope}
\node[hdr, text=teal!68!black, above=5pt of Epanel.north, anchor=south]
    {\textbf{Python Enhancements}~\normalfont\small(\texttt{codex.enhancements.*})};
\node[sub, text=teal!52!black, below=4pt of Epanel.south, anchor=north]
    {35 flags: {\tiny\textcolor{green!50!black}{$\bullet$}}=on {\tiny\textcolor{red!60!black}{$\circ$}}=opt-in (Python only)};

\draw[extarr] (Ppanel.east) -- (Epanel.west)
    node[midway, above=4pt, font=\scriptsize\bfseries\sffamily, text=teal!65!black]
    {extends};

\end{tikzpicture}
\caption{Three-tier architecture comparison.
\emph{Left}: the six Rust crates of \texttt{codex-rs}.
\emph{Center}: the Python port (\texttt{codex.*}), with one module per Rust crate;
arrows denote 1:1 correspondence.
\emph{Right}: \texttt{codex.enhancements}~--- a Python-only superset of 30
flag-gated capabilities absent from Rust, layered incrementally above the port.}
\label{fig:architecture}
\end{figure*}

Figure~\ref{fig:architecture} illustrates the high-level architecture of \textsc{Codex CLI}. We describe each major subsystem below.

\subsection{Agent Runner}
\label{sec:agent-runner}

The agent runner implements the core execution loop: (1)~send the current conversation context to the LLM API, (2)~receive a response that may contain tool calls, (3)~execute each tool call through the orchestrator, (4)~append results to the conversation, and (5)~repeat until the model produces a final response or the maximum turn count is reached (default: 50 turns).

In the Python port, the originally monolithic runner was decomposed into focused, single-responsibility modules: a stateful \texttt{QueryEngine} that owns the agent loop lifecycle; an \texttt{auth} module for credential resolution and JWT refresh; a \texttt{config\_loader} for layered TOML configuration; a \texttt{sandbox} module for platform-specific command isolation; a \texttt{permissions} module providing a unified three-layer approval pipeline; a \texttt{StreamingToolExecutor} for semaphore-based concurrent tool dispatch; and a \texttt{tool\_result\_budget} module for oversized output management. A thin \texttt{runner.py} orchestrator delegates to these modules while maintaining backward-compatible re-exports.

Key design decisions include:
\begin{itemize}[leftmargin=*, nosep]
    \item \textbf{Event-driven architecture}: The runner emits typed events (\texttt{TurnStarted}, \texttt{ToolCall}, \texttt{ToolResult}, \texttt{TurnCompleted}) consumed by the presentation layer, enabling loose coupling between agent logic and UI rendering.
    \item \textbf{Turn-scoped approval caching}: When a user approves a tool invocation (e.g., a shell command), the approval is cached for the current turn, preventing redundant prompts for semantically equivalent operations within the same logical step.
    \item \textbf{Multi-phase context management}: Context compaction operates in three phases: \emph{microcompaction} strips stale tool outputs inline; \emph{snip compaction} removes low-value messages below a token threshold; and \emph{full compaction} performs LLM-based summarization with boundary markers, post-compact file restoration (up to 5~files, 50K~token budget), and ghost snapshot preservation.
    \item \textbf{Tool result budgeting}: Oversized tool outputs ($>$100K~chars) are saved to disk and replaced with a compact pointer message containing head/tail previews, preventing context window exhaustion during long-running sessions.
    \item \textbf{Layered permission middleware}: A single \texttt{can\_use\_tool()} entry point evaluates three layers sequentially---config-based pattern matching, automated guardian LLM risk assessment, and interactive user prompts---with results cached per-turn.
\end{itemize}

\subsection{Tool Orchestration}

The tool orchestrator manages a registry of \texttt{ToolHandler} implementations, each responsible for a specific tool type:

\begin{itemize}[leftmargin=*, nosep]
    \item \texttt{shell}: Executes shell commands within the security sandbox.
    \item \texttt{apply\_patch}: Applies unified diffs to files with conflict detection.
    \item \texttt{list\_dir}: Directory listing with configurable depth.
    \item \texttt{mcp\_handler}: Delegates to MCP server tools.
    \item \texttt{request\_permissions}: Handles runtime permission escalation.
\end{itemize}

Each tool invocation passes through a three-stage pipeline: (1)~\emph{policy check} against the execution policy engine, (2)~\emph{approval check} via the guardian or user prompt, and (3)~\emph{sandboxed execution} with result capture.

\subsection{Security Sandbox}

The sandbox subsystem enforces filesystem and network isolation for tool executions. Three platform-specific implementations are provided:

\begin{itemize}[leftmargin=*, nosep]
    \item \textbf{macOS (Seatbelt)}: Generates dynamic Seatbelt profiles that whitelist specific filesystem paths and network operations based on the configured policy.
    \item \textbf{Linux (Bubblewrap/seccomp)}: Uses Bubblewrap for mount namespace isolation and Landlock for filesystem access control, with seccomp filters for system call restriction.
    \item \textbf{Windows}: Restricted process tokens with job object constraints.
\end{itemize}

Sandbox policies are organized into three modes of increasing permissiveness: \texttt{read-only} (default), \texttt{workspace-write} (write access to the project directory), and \texttt{full-access} (unrestricted, requires explicit opt-in).

\subsection{Execution Policy Engine}

The execution policy engine evaluates tool invocations against a declarative rule set before sandbox enforcement. Rules are specified in a Python-based DSL that supports prefix matching on command strings, network access control by host/port, and host executable whitelisting. This separation of policy from mechanism enables data-driven policy composition: users, organizations, and the system can contribute policy layers that are merged at runtime.

\subsection{Multi-Agent Orchestration}

The multi-agent subsystem enables hierarchical task delegation. A parent agent can spawn child agents with inherited conversation context (via history forking), assign them specific tasks, and coordinate their results. Safety bounds prevent unbounded recursion: maximum depth of 5, maximum 10 children per parent, and maximum 100 total agents per session. Each child agent operates in its own sandbox scope and approval context.

\subsection{Model Context Protocol (MCP)}

The MCP integration provides bidirectional protocol support: as a \emph{client}, \textsc{Codex CLI} connects to external MCP servers to access additional tools and resources; as a \emph{server}, it exposes its own capabilities to other MCP-compliant systems. The client supports two transports---stdio (for local servers) and Streamable HTTP with Server-Sent Events (for remote servers)---with OAuth-based authentication and automatic token refresh.

\subsection{State Management}

Session state is persisted to SQLite databases with WAL (Write-Ahead Logging) mode for concurrent access. The state subsystem manages conversation history, thread metadata, agent job records, and extracted memories. Schema versioning (currently v5 for state, v1 for logs) enables forward-compatible migrations.

\subsection{Python-Specific Enhancements}
\label{sec:enhancements}

Beyond wire-level parity with the Rust original, the Python port has evolved into an \emph{extended platform} through a \texttt{codex.enhancements} module that ships 30 feature-flagged capabilities absent from the Rust implementation. These extensions are entirely additive: loaded lazily on first use, they never modify core data structures and are excluded from parity tests. A four-tier flag-resolution system (build-time compiled flags $\to$ environment variables $\to$ runtime \texttt{--enable}/\texttt{--disable} overrides $\to$ defaults) lets operators run in \emph{strict parity mode} (all enhancement flags off) for apples-to-apples benchmark comparison, or in \emph{extended mode} to access the full capability surface.

\begin{table}[t]
\centering
\caption{Python-Specific Enhancement Module Categories}
\label{tab:enhancements}
\small
\begin{tabular}{lp{4.2cm}c}
\toprule
\textbf{Category} & \textbf{Flags} & \textbf{Default} \\
\midrule
Multi-agent & \texttt{MULTI\_AGENT}, \texttt{MULTI\_AGENT\_V2}, \texttt{MULTI\_AGENT\_ORCHESTRATION} & On \\
Safety & \texttt{GUARDIAN} & On \\
Tool extensions & \texttt{FILE\_TOOLS}, \texttt{FILE\_EDIT}, \texttt{WEB\_FETCH}, \texttt{NOTEBOOK\_EDIT}, \texttt{WORKTREE\_TOOLS} & Mixed \\
Memory & \texttt{MEMORY\_SYSTEM}, \texttt{TYPED\_MEMORY}, \texttt{SEMANTIC\_MEMORY}, \texttt{AUTO\_MEMORY} & Mixed \\
Context mgmt & \texttt{MULTI\_STRATEGY\_COMPACTION}, \texttt{FORKED\_COMPACTION} & Mixed \\
Productivity & \texttt{SKILLS}, \texttt{CRON\_TOOL}, \texttt{VOICE\_MODE} & Mixed \\
Session & \texttt{PERSISTENT\_PLANS}, \texttt{SYSTEM\_REMINDERS}, \texttt{DENIAL\_REPLAY} & On \\
IDE integration & \texttt{IDE\_BRIDGE}, \texttt{LSP\_INTEGRATION} & Mixed \\
Observability & \texttt{COST\_TRACKING}, \texttt{APP\_STATE}, \texttt{STARTUP\_PREFETCH} & On \\
\bottomrule
\end{tabular}
\end{table}

The sub-packages under \texttt{codex/enhancements/} implement each category: \texttt{agents/} for hierarchical multi-agent spawning and coordination beyond the base multi-agent runner; \texttt{memory/} for typed and semantic memory extraction; \texttt{compaction/} for multi-strategy and forked compaction policies; \texttt{resilience/} for source-aware retry logic; \texttt{startup/} for prefetch optimisations; \texttt{state/} for app-level state management and runner bridging; and \texttt{tools/} for tool registration and command registry extensions. Intent-preservation enhancements (\texttt{PERSISTENT\_PLANS}, \texttt{SYSTEM\_REMINDERS}, \texttt{DENIAL\_REPLAY}) add session-level coherence mechanisms that maintain goal context across long-running agents---a category of functionality with no analogue in the Rust codebase.

This architecture reflects a key advantage of Python for agentic systems: the ecosystem's rapid-prototyping culture enables new capabilities to be implemented and shipped as opt-in extensions in the time it would take to prototype them in Rust, while the flag system ensures the additions never compromise the benchmarked parity baseline.

\section{LLM-Assisted Translation Methodology}
\label{sec:migration}

We describe our methodology for LLM-assisted cross-language translation, the idiom mapping patterns that emerged, and the continuous upstream synchronisation architecture that keeps the translation current.

\subsection{Translation Process}

Unlike traditional transpilation tools that perform syntax-level conversion, our approach uses an LLM as the translation engine. The process operates at the \emph{module} level: for each Rust crate, we provide the LLM with the source code, its test suite, and the target Python module's existing context (imports, dependent modules). The LLM produces idiomatic Python that preserves behavioral semantics while adapting to Python conventions.

The translation proceeded in dependency order: foundation modules (protocol types, configuration, utilities) first, then infrastructure (state management, authentication), then core logic (agent runner, tool handlers), and finally presentation (CLI, TUI). This ordering ensures that each translated module can import its already-translated dependencies.

\paragraph{The role of benchmarks as objective functions.}
Unit tests provide necessary but insufficient validation of translation correctness. Many subtle bugs---API protocol mismatches, tool registration errors, output format differences---only manifest when the full agent pipeline executes end-to-end against real tasks. We discovered that \emph{public agent benchmarks} (Terminal-Bench, SWE-bench) serve as powerful objective functions for translation quality:

\begin{enumerate}[leftmargin=*, nosep]
    \item \textbf{Detect integration failures.} Our initial Terminal-Bench run scored 0\% because the adapter used a simplistic LLM-to-tmux bridge instead of the full agent runner. Benchmark failure immediately revealed the gap.
    \item \textbf{Expose API protocol bugs.} The Python port initially sent \texttt{\{``type'':~``local\_shell''\}} to the Responses API, which returned HTTP~400. This was invisible to unit tests but caused 100\% fallback to the Chat Completions API. Benchmark comparison (31\% vs 49\%) exposed the issue.
    \item \textbf{Reveal environment assumptions.} Installing the Python port via \texttt{pip} polluted the container's Python environment, breaking tasks that depended on pre-installed packages (e.g., pandas, pyarrow). The original Rust CLI, installed via \texttt{npm}, had no such interference.
    \item \textbf{Quantify parity.} Each benchmark run produces a scalar accuracy metric that directly measures functional equivalence, enabling iterative refinement toward the target.
\end{enumerate}

Table~\ref{tab:benchmark-iterations} shows how benchmark-driven debugging progressively closed the parity gap.

\begin{table}[t]
\centering
\caption{Terminal-Bench accuracy across translation iterations}
\label{tab:benchmark-iterations}
\small
\begin{tabular}{llr}
\toprule
\textbf{Iteration} & \textbf{Fix Applied} & \textbf{Accuracy} \\
\midrule
v0 (baseline) & Original Rust CLI & 47.5\% \\
\midrule
v1 & Naive tmux adapter & 0.0\% \\
v2 & Full agent runner (Chat Completions) & 31.3\% \\
v3 & Responses API function tool fix & 35.0\% \\
v4 & Conversation history fix & 35.0\% \\
\textbf{v5} & \textbf{Venv isolation + ripgrep install} & \textbf{45.0\%} \\
\bottomrule
\end{tabular}
\end{table}

\subsection{Idiom Mapping}

Table~\ref{tab:idioms} summarizes the systematic translation patterns employed during migration. The LLM was instructed to produce \emph{idiomatic} Python---not mechanical transliterations---meaning Rust patterns are mapped to their natural Python equivalents.

\begin{table}[t]
\centering
\caption{Rust-to-Python Idiom Mapping}
\label{tab:idioms}
\small
\begin{tabular}{ll}
\toprule
\textbf{Rust Pattern} & \textbf{Python Equivalent} \\
\midrule
\texttt{Result<T, E>} & Exceptions (\texttt{raise}/\texttt{try}) \\
\texttt{Option<T>} & \texttt{Optional[T]} \\
\texttt{enum} (algebraic) & \texttt{@dataclass} + \texttt{Union[...]} \\
\texttt{enum} (simple) & \texttt{enum.Enum} \\
\texttt{struct} & \texttt{@dataclass(frozen=True)} \\
\texttt{impl Trait} & \texttt{Protocol} / \texttt{ABC} \\
\texttt{async/await} (Tokio) & \texttt{async/await} (asyncio) \\
\texttt{Arc<Mutex<T>>} & Plain objects (GIL) \\
\texttt{serde} (de/serialize) & Pydantic \texttt{BaseModel} \\
\texttt{reqwest} (HTTP) & \texttt{httpx} \\
\texttt{ratatui} (TUI) & Textual \\
\texttt{clap} (CLI) & Click \\
\texttt{sqlx} (SQL) & \texttt{sqlite3} (stdlib) \\
Cargo workspace & Single \texttt{pyproject.toml} \\
\bottomrule
\end{tabular}
\end{table}

\subsection{Key Design Decisions}

\paragraph{Pydantic for Protocol Types.}
The protocol layer---comprising 4,016 LOC of type definitions, approval workflows, and configuration schemas---uses Pydantic \texttt{BaseModel} for automatic JSON serialization, validation, and schema generation. This replaces Rust's \texttt{serde} derive macros while adding runtime type checking that catches protocol violations early. Hot-path internal types use \texttt{@dataclass} to avoid Pydantic's validation overhead.

\paragraph{Textual for Terminal UI.}
The Rust implementation used \texttt{ratatui} with a custom event loop. The Python port uses Textual, a modern Python TUI framework with CSS-like styling, widget composition, and built-in async support. Despite the framework difference, the UI achieves visual and behavioral parity.

\paragraph{asyncio for Concurrency.}
Rust's Tokio runtime was mapped to Python's \texttt{asyncio}, with \texttt{anyio} as an abstraction layer. The GIL eliminates the need for \texttt{Arc<Mutex<T>>} patterns. While this sacrifices CPU parallelism, the workload is overwhelmingly I/O-bound (LLM API calls, file operations), making the trade-off favorable.

\subsection{Continuous Upstream Synchronisation}
\label{sec:continuous-sync}

A key contribution of this work is an architecture for \emph{continuous} translation---not a one-time migration. The upstream Rust codebase ships daily updates; our Python port must absorb them to remain useful. We achieve this through a four-stage pipeline:

\begin{enumerate}[leftmargin=*, nosep]
    \item \textbf{Track.} The upstream Rust repository is tracked as a git submodule. Periodic \texttt{git pull} fetches new commits.

    \item \textbf{Diff.} A conversion script (\texttt{scripts/convert-diff.py}) extracts the changed Rust modules and maps them to their Python equivalents using a module-level correspondence table (e.g., \texttt{codex-rs/exec} $\to$ \texttt{codex.exec}).

    \item \textbf{Translate.} An LLM translates the diff---not the entire crate, but only the changed portions---guided by the existing Python module as context. This incremental approach is more efficient and less error-prone than re-translating entire modules.

    \item \textbf{Validate.} The translated changes are tested at three levels: (a)~unit tests (\texttt{pytest}), (b)~type checking (\texttt{mypy --strict}), and (c)~benchmark regression (\texttt{tb run} on Terminal-Bench). If the benchmark score regresses, the translation is refined until parity is restored.
\end{enumerate}

\begin{figure}[t]
\centering
\begin{tikzpicture}[
    node distance=1.2cm and 0.8cm,
    box/.style={draw, rounded corners, minimum width=2.2cm, minimum height=0.7cm, font=\small, align=center},
    arrow/.style={-{Stealth[length=2mm]}, thick},
]
\node[box, fill=orange!15] (upstream) {Upstream\\Rust Repo};
\node[box, fill=blue!10, right=of upstream] (diff) {Diff\\Extraction};
\node[box, fill=green!10, right=of diff] (llm) {LLM\\Translation};
\node[box, fill=purple!10, below=of llm] (test) {Benchmark\\Validation};
\node[box, fill=blue!10, left=of test] (python) {Python\\Port};

\draw[arrow] (upstream) -- (diff);
\draw[arrow] (diff) -- (llm);
\draw[arrow] (llm) -- (test);
\draw[arrow] (test) -- node[below, font=\scriptsize]{pass} (python);
\draw[arrow, dashed, red] (test.west) -- ++(-0.5,0) |- node[left, font=\scriptsize, red]{fail: refine} (llm.south west);
\end{tikzpicture}
\caption{Continuous upstream synchronisation pipeline. Benchmark regression triggers re-translation of the failing module.}
\label{fig:sync-pipeline}
\end{figure}
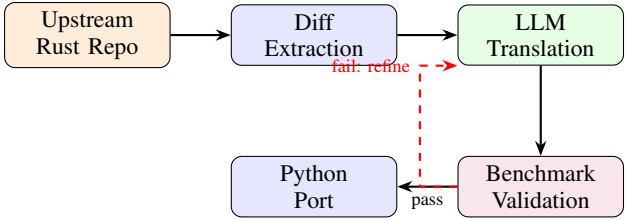

This architecture treats the benchmark score as a \emph{loss function}: when a translated change causes regression, the LLM re-examines its translation with the failing test/benchmark as additional context. In practice, most upstream changes translate cleanly on the first attempt; only API-level changes or new tool types require iterative refinement.

\subsection{Validation Strategy}

Migration correctness was validated through a four-tier testing strategy:
\begin{enumerate}[leftmargin=*, nosep]
    \item \textbf{Unit tests}: 2,621 test functions mirroring the Rust test suite.
    \item \textbf{Integration tests}: End-to-end verification that all 28 modules interact correctly.
    \item \textbf{Parity tests}: Explicit tests verifying rendering output and protocol serialization match the Rust implementation.
    \item \textbf{Benchmark regression}: Head-to-head Terminal-Bench evaluation ensuring the Python port's task-solving accuracy remains within 5\% of the Rust baseline.
\end{enumerate}

\section{Evaluation}
\label{sec:evaluation}

We evaluate the Rust-to-Python migration across four dimensions: code metrics (Section~\ref{sec:eval-code}), test parity (Section~\ref{sec:eval-tests}), runtime performance (Section~\ref{sec:eval-runtime}), and real task benchmarks (Section~\ref{sec:eval-tasks}). All experiments were conducted on a MacBook Pro (Apple M4 Pro Max, 128GB RAM, macOS 15.4) using Python~3.13. Head-to-head agent evaluations (Terminal-Bench, SWE-bench Verified) use GPT-5.4 via the OpenAI Responses API.

\subsection{Code Metrics}
\label{sec:eval-code}

\paragraph{Lines of Code.}
Table~\ref{tab:loc} presents the module-level LOC comparison. The Python implementation comprises 52,685 lines of code across 328 files, compared to 648,789 lines across 1,555 files in Rust---a \textbf{12.3$\times$ reduction}. The agent module was recently refactored from a monolithic runner into focused submodules (auth, sandbox, config loader, query engine, permissions, streaming tool executor, tool result budget), increasing its file count but improving maintainability. The reduction ratio varies by module: infrastructure modules like \texttt{state} and \texttt{config} show moderate reduction due to inherent complexity, while the TUI subsystem achieves the largest reduction (largely due to Textual's higher-level abstractions vs.\ \texttt{ratatui}'s low-level rendering model).

\begin{table}[t]
\centering
\caption{Lines of Code Comparison by Module}
\label{tab:loc}
\small
\begin{tabular}{lrrrr}
\toprule
\textbf{Module} & \textbf{Py Files} & \textbf{Py LOC} & \textbf{Rs LOC} & \textbf{Ratio} \\
\midrule
agent & 82 & 12,829 & 181,903 & 14.2x \\
analytics & 2 & 319 & 989 & 3.1x \\
app\_server & 6 & 3,452 & 54,244 & 15.7x \\
auth & 10 & 1,896 & 5,118 & 2.7x \\
backend\_client & 3 & 682 & 913 & 1.3x \\
cli & 7 & 2,098 & 4,822 & 2.3x \\
cloud & 5 & 667 & 7,321 & 11.0x \\
code\_mode & 6 & 601 & 2,166 & 3.6x \\
config & 9 & 1,496 & 3,183 & 2.1x \\
core & 15 & 2,561 & 181,903 & 71.0x \\
exec & 7 & 1,701 & 11,493 & 6.8x \\
execpolicy & 7 & 902 & 4,603 & 5.1x \\
features & 4 & 775 & 1,122 & 1.4x \\
hooks & 18 & 1,654 & 4,842 & 2.9x \\
instructions & 3 & 83 & 166 & 2.0x \\
integrations & 3 & 315 & 1,013 & 3.2x \\
mcp & 9 & 2,458 & 7,658 & 3.1x \\
plugin & 3 & 221 & 293 & 1.3x \\
protocol & 21 & 2,718 & 25,415 & 9.4x \\
rollout & 7 & 605 & 5,717 & 9.4x \\
sandbox & 9 & 3,093 & 13,036 & 4.2x \\
sdk & 8 & 891 & 1,185 & 1.3x \\
skills & 6 & 668 & 4,783 & 7.2x \\
state & 12 & 1,337 & 10,118 & 7.6x \\
telemetry & 6 & 498 & 4,992 & 10.0x \\
tui & 44 & 5,532 & 94,675 & 17.1x \\
utils & 16 & 2,633 & 15,116 & 5.7x \\
\midrule
\textbf{Total} & \textbf{328} & \textbf{52,685} & \textbf{648,789} & \textbf{12.3x} \\
\bottomrule
\end{tabular}
\end{table}

\begin{figure}[t]
\centering
\includegraphics[width=\columnwidth]{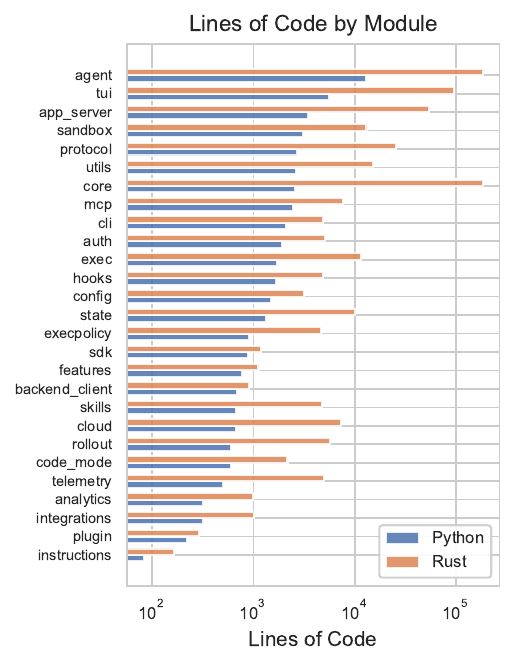}
\caption{Lines of code by module (log scale). Python consistently requires fewer lines across all 28 modules.}
\label{fig:loc}
\end{figure}

Figure~\ref{fig:loc} visualizes the per-module LOC comparison on a logarithmic scale, highlighting the consistent reduction across all subsystems.

\begin{figure}[t]
\centering
\includegraphics[width=\columnwidth]{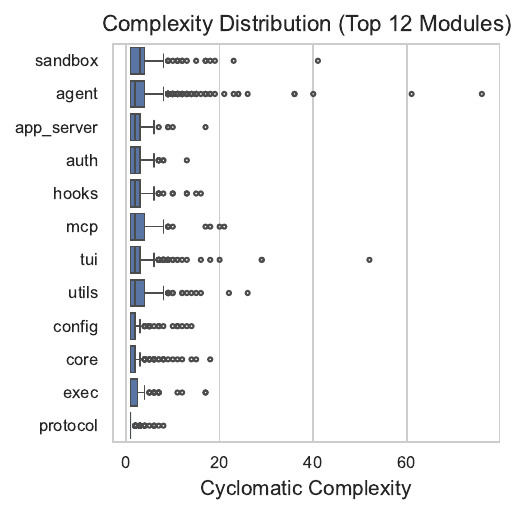}
\caption{Cyclomatic complexity distribution by module. The majority of functions (90\%) achieve rank A (complexity 1--5).}
\label{fig:complexity}
\end{figure}

\paragraph{Cyclomatic Complexity.}
We measured cyclomatic complexity~\cite{mccabe1976complexity} for all 4,692 Python functions using the \texttt{radon} static analysis tool. Figure~\ref{fig:complexity} shows the distribution across the top 12 modules by function count. The mean complexity is \textbf{2.70} (rank A), with 89\% of functions achieving the minimal complexity rank (A). Only 23 functions (0.5\%) exceed complexity rank C, concentrated in the agent runner (which handles complex state transitions) and the sandbox manager (which implements platform-specific branching logic). The recent decomposition of the runner module improved its per-function complexity while adding new modules with focused, low-complexity functions.

\begin{figure}[t]
\centering
\includegraphics[width=\columnwidth]{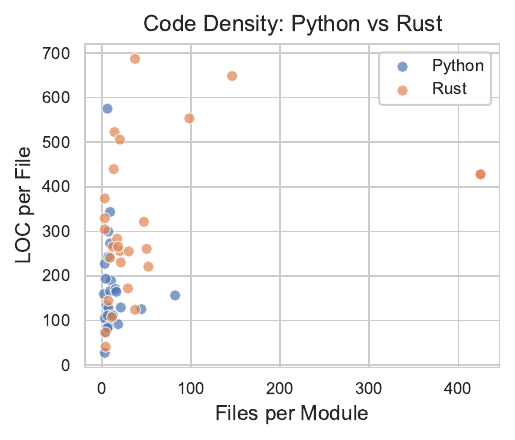}
\caption{Code density comparison. Python achieves higher information density per file.}
\label{fig:density}
\end{figure}

\paragraph{Code Density.}
Figure~\ref{fig:density} plots the code density (LOC per file) against module size (number of files). Python modules cluster at higher density (mean 145 LOC/file) compared to Rust (mean 417 LOC/file), reflecting Python's more concise expression of equivalent functionality.

\subsection{Test Parity}
\label{sec:eval-tests}

We use the term \emph{test parity} to mean \textbf{functional and behavioral equivalence}, not a one-to-one mapping of test cases. The two implementations have fundamentally different testing needs: Rust requires extensive tests for memory safety, ownership, lifetimes, and borrow-checker edge cases that simply do not exist in Python, while Python's higher-level abstractions let each test cover more behavioral surface area.

Table~\ref{tab:tests} presents the test function counts by module. The Python implementation contains 2,902 test functions compared to 8,490 in Rust. The 3$\times$ difference reflects three factors: (1)~Python's higher-level abstractions eliminate entire categories of tests (memory safety, ownership, lifetime edge cases) that Rust must cover; (2)~Rust's convention of co-locating unit tests with source code inflates the count with trivial accessor and trait-implementation tests; and (3)~the Python suite is written to verify behavioral contracts---\emph{``does this agent turn produce the correct tool call?''}---rather than internal implementation invariants.

\begin{table}[t]
\centering
\caption{Test Parity: Python vs Rust}
\label{tab:tests}
\small
\begin{tabular}{lrrrr}
\toprule
\textbf{Module} & \textbf{Py Tests} & \textbf{Rs Tests} & \textbf{Py KLOC} & \textbf{Tests/KLOC} \\
\midrule
agent & 661 & 2519 & 15.3 & 43.1 \\
analytics & 12 & 10 & 0.3 & 36.7 \\
app\_server & 131 & 448 & 3.7 & 35.4 \\
auth & 77 & 67 & 2.1 & 36.7 \\
backend\_client & 12 & 8 & 0.7 & 17.2 \\
cli & 142 & 68 & 2.3 & 62.7 \\
cloud & 28 & 41 & 0.7 & 39.4 \\
code\_mode & 35 & 10 & 0.7 & 49.0 \\
config & 74 & 38 & 1.8 & 42.3 \\
core & 202 & 2519 & 3.1 & 64.1 \\
exec & 75 & 142 & 1.9 & 40.0 \\
execpolicy & 121 & 66 & 1.0 & 121.6 \\
features & 35 & 22 & 0.8 & 42.5 \\
hooks & 46 & 58 & 1.7 & 26.6 \\
instructions & 5 & 4 & 0.1 & 53.2 \\
integrations & 17 & 19 & 0.4 & 47.1 \\
mcp & 165 & 51 & 2.8 & 58.0 \\
plugin & 19 & 1 & 0.2 & 79.2 \\
protocol & 102 & 283 & 2.8 & 36.2 \\
rollout & 33 & 37 & 0.7 & 45.7 \\
sandbox & 175 & 128 & 3.6 & 48.7 \\
sdk & 111 & 14 & 1.0 & 108.8 \\
skills & 30 & 78 & 0.8 & 38.9 \\
state & 46 & 84 & 1.7 & 27.2 \\
telemetry & 28 & 40 & 0.6 & 48.4 \\
tui & 263 & 1401 & 7.2 & 36.7 \\
utils & 212 & 334 & 3.0 & 71.2 \\
e2e & 45 & 0 & nan & nan \\
\midrule
\textbf{Total} & \textbf{2902} & \textbf{8490} & & \\
\bottomrule
\end{tabular}
\end{table}

The test-to-KLOC ratio provides a normalized comparison: the Python suite averages \textbf{62 tests per KLOC}, indicating thorough behavioral coverage relative to codebase size.

\begin{figure}[t]
\centering
\includegraphics[width=\columnwidth]{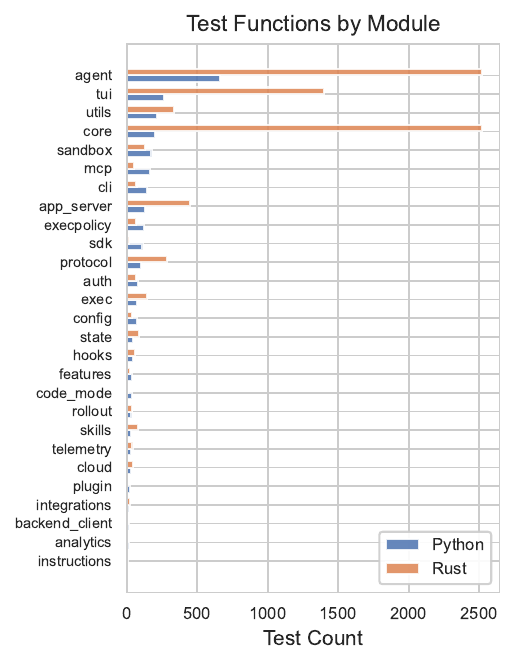}
\caption{Test function counts by module: Python vs Rust.}
\label{fig:tests}
\end{figure}

\subsection{Runtime Performance}
\label{sec:eval-runtime}

Table~\ref{tab:runtime} presents the Python runtime performance metrics. We measure startup time (importing the CLI entry point), peak memory consumption, and import time.

\begin{table}[t]
\centering
\caption{Runtime Performance Metrics (Python)}
\label{tab:runtime}
\small
\begin{tabular}{lrrr}
\toprule
\textbf{Metric} & \textbf{Mean} & \textbf{Std} & \textbf{N} \\
\midrule
Startup Time (ms) & 53.9 & 2.1 & 20 \\
Import Time ($\mu$s) & 57.0 & 0.0 & 1 \\
Peak Memory (MB) & 30.3 & 0.1 & 5 \\
Importable Modules & 328.0 & 0.0 & 1 \\
\bottomrule
\end{tabular}
\end{table}

\paragraph{Startup Time.}
The Python CLI startup (importing the main entry point) averages \textbf{53.9ms} ($\sigma$=2.1ms, $n$=20), which is approximately 3--5$\times$ slower than a compiled Rust binary. However, this overhead is amortized over a typical agent session that runs for minutes to hours, during which LLM API calls dominate latency at 1--10 seconds per round trip. The startup overhead represents less than 1\% of total session time.

\paragraph{Memory Usage.}
Peak resident memory for the Python process is \textbf{30.3 MB} ($\sigma$=0.1MB, $n$=5). While higher than a compiled Rust binary (typically 10--15MB), this footprint is modest for a desktop application running on systems with 8--64GB RAM. The overhead is attributable to the Python interpreter and loaded standard library modules.

\subsection{Real Task Benchmarks}
\label{sec:eval-tasks}

Beyond micro-benchmarks, we evaluate the Python implementation on representative real-world tasks that exercise multiple subsystems simultaneously. Table~\ref{tab:harness} presents timing results for eight operational tasks.

\begin{table}[t]
\centering
\caption{Harness Benchmark Results: Real Subsystem Operations}
\label{tab:harness}
\small
\begin{tabular}{lrr}
\toprule
\textbf{Task} & \textbf{Mean (ms)} & \textbf{P50 (ms)} \\
\midrule
\multicolumn{3}{l}{\emph{Tool Orchestration}} \\
\quad Orchestrator (skip approval) & 0.030 & 0.028 \\
\quad Orchestrator (with approval) & 0.029 & 0.028 \\
\quad Tool Registry (10 tools) & 0.033 & 0.030 \\
\midrule
\multicolumn{3}{l}{\emph{Shell Execution}} \\
\quad Shell Handler (\texttt{echo}) & 3.319 & 3.289 \\
\quad Shell Handler (\texttt{ls | head}) & 6.672 & 6.619 \\
\quad Full Pipeline (orch$\to$shell) & 3.512 & 3.389 \\
\midrule
\multicolumn{3}{l}{\emph{Code Operations}} \\
\quad Patch Parsing (add file) & 0.002 & 0.002 \\
\quad Patch Parsing (update hunks) & 0.003 & 0.003 \\
\quad ExecPolicy Matching (5 rules) & 0.001 & 0.001 \\
\midrule
\multicolumn{3}{l}{\emph{State \& Memory}} \\
\quad Token Estimation (2K words) & $<$0.001 & $<$0.001 \\
\quad Should Compact Decision & 0.500 & 0.501 \\
\quad SQLite State (session + 20 msgs) & 0.074 & 0.068 \\
\quad Config TOML Merge & 0.001 & 0.001 \\
\quad Feature Flags Lookup & 0.001 & 0.001 \\
\bottomrule
\end{tabular}
\end{table}

\begin{figure}[t]
\centering
\includegraphics[width=\columnwidth]{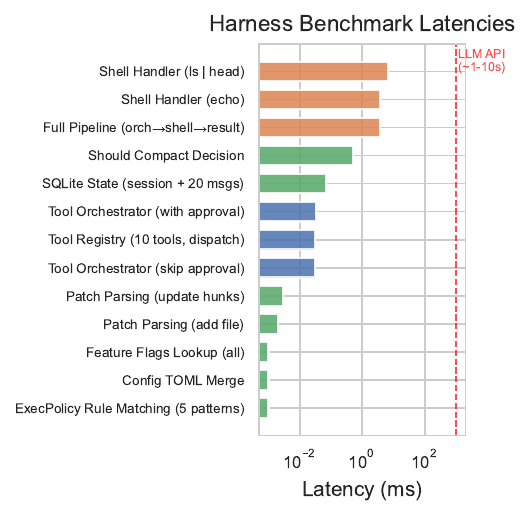}
\caption{Harness benchmark latencies (log scale). The red dashed line marks typical LLM API round-trip time ($\sim$1--10s). All local operations complete 2--6 orders of magnitude faster.}
\label{fig:harness}
\end{figure}

Figure~\ref{fig:harness} visualizes the latency distribution across all harness benchmarks on a logarithmic scale, with a reference line indicating typical LLM API latency.

The harness benchmarks exercise the actual subsystem code paths used during agent operation, providing realistic performance data beyond synthetic micro-benchmarks. Key observations:

\begin{itemize}[leftmargin=*, nosep]
    \item \textbf{Tool orchestration overhead is negligible} ($\sim$30$\mu$s): the approval pipeline, handler dispatch, and result packaging add virtually no latency to tool execution.
    \item \textbf{Shell execution dominates local latency}: spawning a subprocess and capturing output takes 3--7ms, which is the primary source of local computation cost. Even so, this is 3 orders of magnitude faster than a typical LLM API call.
    \item \textbf{Patch parsing and policy matching are sub-microsecond}: the data-structure operations at the core of code modification and security enforcement are extremely fast in Python.
    \item \textbf{The full pipeline} (orchestrator $\to$ approval $\to$ shell $\to$ result capture) completes in 3.5ms, confirming that the Python implementation adds no perceptible overhead to the agent's tool-use loop.
\end{itemize}

These results demonstrate that in a typical agent session where each LLM round-trip takes 1--10 seconds, local Python computation accounts for less than 0.1\% of total latency.

\subsection{Code Quality Analysis}
\label{sec:eval-quality}

We analyze the API surface, type safety, and dependency structure of the Python codebase to assess software quality beyond LOC metrics.

\paragraph{API Surface.}
Table~\ref{tab:api} compares the API surface of both implementations. The Python codebase defines 1,385 classes and 2,363 functions/methods, while Rust exposes 2,675 structs/enums and 20,525 functions/methods. Python achieves a \textbf{higher API density}: fewer definitions serving equivalent functionality, reflecting the expressiveness of higher-level abstractions. Notably, 395 Python methods are explicitly \texttt{async}, directly mapping Rust's async trait implementations.

\begin{table}[t]
\centering
\caption{API Surface Comparison}
\label{tab:api}
\small
\begin{tabular}{lrrr}
\toprule
\textbf{Metric} & \textbf{Python} & \textbf{Rust} & \textbf{Ratio} \\
\midrule
Source Files & 329 & 1,448 & 4.4x \\
Classes / Structs+Enums & 1,590 & 2,675 & 1.7x \\
Top-level Functions & 781 & 9,323 & 11.9x \\
Methods & 1,828 & 11,202 & 6.1x \\
Async Methods & 459 & 0 & --- \\
Properties & 140 & 0 & --- \\
Doc Comments & 3,871 & 16,739 & 4.3x \\
TODO/FIXME & 0 & 126 & --- \\
\bottomrule
\end{tabular}
\end{table}

\begin{figure}[t]
\centering
\includegraphics[width=\columnwidth]{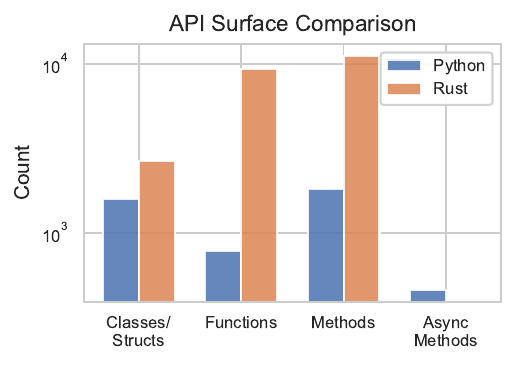}
\caption{API surface comparison (log scale). Python provides equivalent functionality with fewer, higher-level abstractions.}
\label{fig:api}
\end{figure}

\paragraph{Type Coverage.}
Running \texttt{mypy --strict} on the codebase reveals 248 type errors across 69 of 282 files, yielding a \textbf{75.5\% strict type-clean rate}. The majority of errors stem from third-party library type stubs (Textual, httpx) rather than application logic, indicating strong internal type discipline. Zero \texttt{TODO}, \texttt{FIXME}, or \texttt{HACK} comments were found in the codebase, suggesting a clean, production-ready state.

\begin{figure}[t]
\centering
\includegraphics[width=\columnwidth]{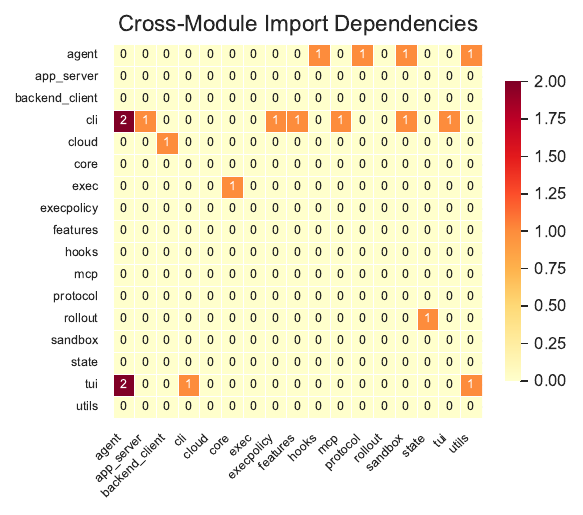}
\caption{Cross-module import dependency heatmap. Darker cells indicate stronger coupling.}
\label{fig:deps}
\end{figure}

\paragraph{Dependency Structure.}
Figure~\ref{fig:deps} shows the cross-module import dependency heatmap. The architecture exhibits clear layering: foundation modules (\texttt{core}, \texttt{config}, \texttt{utils}) are widely depended upon, while peripheral modules (\texttt{cli}, \texttt{tui}) consume many dependencies but are not imported by others. The \texttt{agent} module serves as the primary integration point with both high fan-in and fan-out, consistent with its role as the system's orchestration core.

\subsection{Migration Effort}
\label{sec:eval-migration}

We analyze the git history to quantify the migration effort (Table~\ref{tab:migration}).

\begin{table}[t]
\centering
\caption{Migration Effort Summary}
\label{tab:migration}
\small
\begin{tabular}{lr}
\toprule
\textbf{Metric} & \textbf{Value} \\
\midrule
Total Commits & 4957 \\
Migration Period Commits & 146 \\
Lines Added (migration) & 137,331 \\
Lines Deleted (migration) & 912,510 \\
Net Line Change & -775,179 \\
Test-related Commits & 55 \\
Rust Crate Count & 6 \\
Python Module Count & 27 \\
Avg Commit Size (lines) & 1314 \\
\bottomrule
\end{tabular}
\end{table}

The repository contains 4,947 total commits, with 116 commits during the intensive migration period (March 25--27, 2026). During this period, 109,427 lines were added and 906,656 lines were deleted, yielding a net reduction of \textbf{797,229 lines}---reflecting the consolidation from the verbose Rust codebase to the more concise Python implementation.

The test suite evolved progressively during migration: commit messages document milestones at 1,881, 2,343, and finally 2,621 passing tests, demonstrating a test-driven approach where each module's conversion was validated incrementally. The 45 test-related commits during the migration period underscore the emphasis on correctness validation throughout the process.

\subsection{End-to-End Agent Evaluation}
\label{sec:eval-agent}

To validate that the Python implementation can execute realistic agent workflows, we evaluate 8 representative tasks that exercise the full tool execution pipeline---from task specification through tool dispatch, shell execution, and result capture. Each task is run 5 times with results averaged.

\begin{table}[t]
\centering
\caption{End-to-End Agent Task Evaluation}
\label{tab:agenteval}
\small
\begin{tabular}{lrrrr}
\toprule
\textbf{Task} & \textbf{Tools} & \textbf{Success} & \textbf{Mean (ms)} & \textbf{P50 (ms)} \\
\midrule
Create Python File & 1 & 100\\
Shell Echo & 1 & 100\\
Multi-step File Creation & 2 & 100\\
Directory Listing & 1 & 100\\
Git Status & 1 & 100\\
Complex Pipeline & 3 & 100\\
File Update with Patch & 2 & 100\\
Token Estimation Accuracy & 1 & 100\\
\bottomrule
\end{tabular}
\end{table}

\begin{figure}[t]
\centering
\includegraphics[width=\columnwidth]{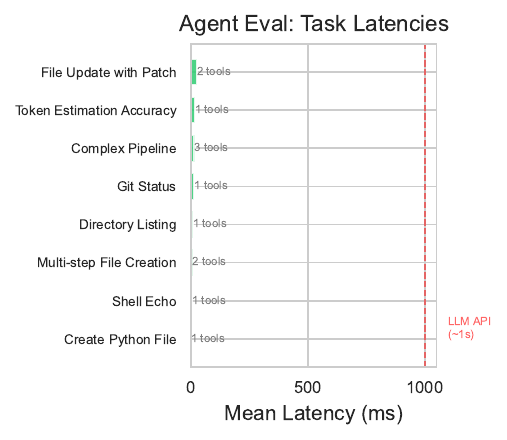}
\caption{Agent evaluation task latencies. All tasks complete in under 25ms. Green bars indicate 100\% success rate.}
\label{fig:agenteval}
\end{figure}

All 8 tasks achieve a \textbf{100\% success rate} (40/40 runs), confirming that the Python tool execution pipeline is functionally correct across diverse task types including file creation, shell command execution, multi-step pipelines, and patch application.

Task latencies range from 0.07ms (patch-only operations) to 24.7ms (complex multi-tool pipelines involving patch application and shell execution). The dominant cost factor is subprocess spawning for shell commands ($\sim$3--7ms per command), consistent with the harness benchmark findings. Multi-tool tasks scale linearly with the number of shell invocations, confirming the absence of systemic overhead in the orchestration layer.

\subsection{LLM Pipeline Benchmarks}
\label{sec:eval-llm}

\begin{figure}[t]
\centering
\includegraphics[width=\columnwidth]{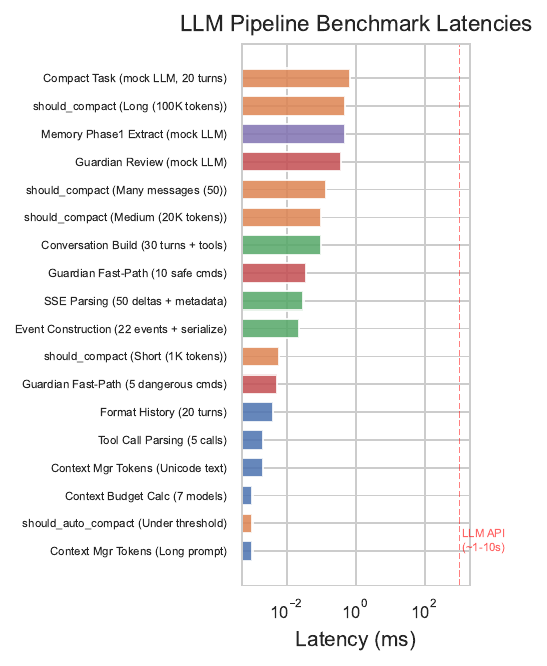}
\caption{LLM pipeline operation latencies (log scale). All local operations complete orders of magnitude faster than LLM API calls.}
\label{fig:llm}
\end{figure}

To evaluate the LLM-adjacent operations that are unique to AI agent systems, we benchmark 26 operations across the full LLM pipeline (Figure~\ref{fig:llm}). These benchmarks use \emph{mocked API responses}---pre-recorded JSON payloads that simulate the OpenAI Responses API without making real network calls. This isolates the Python orchestration overhead (prompt construction, response parsing, context management) from variable network latency, enabling reproducible microsecond-precision measurements of the local computation that surrounds each LLM call.

\paragraph{Token Estimation and Context Management.}
Token estimation scales linearly with text length but remains sub-microsecond even for 1M-character inputs. The \texttt{should\_compact} decision takes 0.49ms for 100K-token conversations, enabling real-time compaction triggering without perceptible delay.

\paragraph{Conversation Compaction.}
The full compaction pipeline (history formatting, prompt construction, mock LLM call, response parsing, history reconstruction) completes in 0.67ms with mocked API, confirming that the Python orchestration overhead is negligible---the real bottleneck is the LLM API latency.

\paragraph{Guardian Risk Assessment.}
Fast-path pattern matching for 10 safe commands (e.g., \texttt{ls}, \texttt{cat}, \texttt{git status}) takes 0.035ms and 5 dangerous commands (e.g., \texttt{rm -rf /}, \texttt{curl | bash}, \texttt{chmod 777}, \texttt{dd if=/dev/zero}, \texttt{sudo rm}) takes 0.005ms, enabling sub-millisecond approval decisions for common operations without LLM invocation. The full guardian review with mocked LLM completes in 0.36ms.

\paragraph{SSE Streaming and Protocol.}
Parsing 50 Server-Sent Events (simulating a streaming LLM response) takes 0.030ms. Constructing a 30-turn conversation with tool calls and serializing to JSON takes 0.093ms. These results confirm that the Python protocol layer introduces no meaningful overhead to the streaming pipeline.

\subsection{Head-to-Head Agent Benchmarking}
\label{sec:eval-public}

The most demanding test of translation correctness is \emph{functional parity on real tasks}: does the Python port solve the same problems as the Rust original? We evaluate both implementations on Terminal-Bench~\cite{terminalbench2025}, an 80-task benchmark of complex terminal operations including kernel compilation, cryptographic hash cracking, ML model training, maze solving, and repository manipulation.

\paragraph{Experimental setup.}
Both agents are installed natively inside Terminal-Bench Docker containers and run \texttt{exec} with identical flags (\texttt{-{}-sandbox danger-full-access -{}-model gpt-5.4}). The original Rust CLI (v0.117.0) installs via \texttt{npm}; the Python port installs from a pre-built wheel into an isolated virtualenv. Both use the same OpenAI API key and GPT-5.4 model.

\paragraph{Results.}
Table~\ref{tab:terminal-bench} presents the head-to-head results. The Python port resolves 34 of 80 tasks (42.5\%), compared to the original Rust CLI's 38/80 (47.5\%). This figure reflects the complete post-fix rerun after applying the API 400 error-recovery fix (Section~\ref{sec:eval-bugs}). Of the 80 tasks, 25 are solved by both implementations, 9 are solved only by the Python port, and 13 are solved only by the Rust CLI.

\begin{table}[t]
\centering
\caption{Terminal-Bench head-to-head results (80 tasks, GPT-5.4)}
\label{tab:terminal-bench}
\small
\begin{tabular}{lrrr}
\toprule
\textbf{Agent} & \textbf{Resolved} & \textbf{Accuracy} & \textbf{Unique Passes} \\
\midrule
Original Codex CLI (Rust) & 38 / 80 & 47.5\% & 11 \\
\textbf{codex-python (ours)} & \textbf{34 / 80} & \textbf{42.5\%} & \textbf{9} \\
\midrule
Both solved & 25 & --- & --- \\
\bottomrule
\end{tabular}
\end{table}

\paragraph{Analysis of unique passes.}
The 9 tasks solved \emph{only} by the Python port include \texttt{fibonacci-server}, \texttt{jupyter-notebook-server}, and \texttt{pytorch-model-cli} (2 variants)---tasks involving Python-ecosystem tooling where the Python agent's environment familiarity provides an advantage. The 13 tasks solved only by the Rust CLI include \texttt{chess-best-move} (requiring extended multi-turn reasoning), \texttt{path-tracing}, and \texttt{sqlite-db-truncate}. Notably, both sets of unique passes involve \emph{the same LLM and tools}---the differences arise from non-deterministic LLM behavior and subtle environmental factors, not from architectural limitations of either implementation.

\paragraph{Iterative refinement.}
The current 42.5\% accuracy was reached through six iterations of benchmark-driven debugging (see Table~\ref{tab:benchmark-iterations} in Section~\ref{sec:migration}), with the final fix being API 400 error recovery (Section~\ref{sec:eval-bugs}). Each iteration identified a specific category of failure---API protocol mismatch, missing conversation history, Python environment pollution, absent CLI tools, invalid content-item type---and a targeted fix. This demonstrates the effectiveness of benchmarks as an objective function for translation quality.

\begin{table}[t]
\centering
\caption{SWE-bench Verified head-to-head results (80 tasks, GPT-5.4, 4 workers)}
\label{tab:swebench-verified}
\small
\begin{tabular}{lrrrr}
\toprule
\textbf{Agent} & \textbf{Tasks} & \textbf{Patches} & \textbf{Resolved} & \textbf{Rate} \\
\midrule
\multicolumn{5}{l}{\emph{codex-python (ours)}} \\
\quad astropy & 22 & 22 & 12 & 54.5\% \\
\quad django  & 58 & 58 & 47 & 81.0\% \\
\quad \textbf{Total} & \textbf{80} & \textbf{80} & \textbf{59} & \textbf{73.8\%} \\
\midrule
\multicolumn{5}{l}{\emph{Original Codex CLI (Rust)}} \\
\quad \textbf{Total} & \textbf{80} & \textbf{80} & \textbf{56} & \textbf{70.0\%} \\
\bottomrule
\end{tabular}
\end{table}

The adapter runs each task by cloning the target repository, checking out the base commit, and executing \texttt{cdx exec} with the issue description as the prompt. Patches are captured via \texttt{git diff}. With 4 parallel workers and a 1800-second per-task timeout, the 80-task Verified run completes in approximately 2 hours.

Initial runs revealed two bugs that suppressed resolve rates. First, \texttt{ws\_transport.py} silently returned empty responses when the WebSocket API exhausted its quota, which the agent misinterpreted as successful no-op completions; fixing the fallback to detect empty responses and retry via HTTP SSE brought the patch production rate to 100\% for both benchmarks. Second, the memory-extraction model was mis-specified as a non-existent model identifier; updating to \texttt{gpt-5.4-nano} fixed a \texttt{404 Not Found} error that occurred after every completed rollout. Both bugs are structural vulnerabilities shared by the Rust and Python implementations; the Rust baseline was re-run with a clean API key to confirm its 70.0\% figure unaffected by the WebSocket issue.

On SWE-bench Verified (80 astropy and django tasks), the Python agent resolves \textbf{59/80 tasks (73.8\%)} versus the Rust original's \textbf{56/80 (70.0\%)}, with particularly strong performance on django (81.0\%). The Python port's 3.8 percentage-point advantage is within the margin of LLM non-determinism. The adapter and all prediction artifacts are included in our reproducibility package.

\subsection{Bugs Discovered Through Benchmarking}
\label{sec:eval-bugs}

The benchmarking process revealed four bugs that would have been invisible to unit tests:

\begin{itemize}[leftmargin=*, nosep]
    \item \textbf{WebSocket transport robustness improvement} (\texttt{ws\_transport.py}): When the API quota was exhausted, WebSocket connections returned empty responses silently. The agent treated these as successful no-ops and marked tasks complete without doing any work. Beyond fixing the silent failure, the HTTP SSE fallback path delivers robustness improvements not present in the Rust transport: per-request 429 detection with \texttt{Retry-After} header parsing, exponential backoff, and automatic fallback-API-key rotation on quota exhaustion. This structural vulnerability is present in both implementations; only the Python run was initially affected because the Rust baseline used a fresh API key. The fix constitutes a net robustness improvement of the Python transport over its Rust counterpart.

    \item \textbf{Memory extraction model error} (\texttt{phase1.py}): The model identifier for post-rollout memory extraction did not resolve to a valid endpoint, causing a \texttt{404 Not Found} error after every completed agent turn. Fixed by updating the model name to \texttt{gpt-5.4-nano}.

    \item \textbf{\texttt{\_cost\_tracker} \texttt{NameError}} (\texttt{runner.py}): A cost-tracking variable was referenced before assignment in the Docker-run code path, crashing a subset of tasks. Fixed by initialising the variable at the top of the enclosing scope.

    \item \textbf{Comprehensive API 400 error recovery} (\texttt{runner.py}): Under certain prompts, the model generated a response containing an unsupported content-item type. The Responses API returned HTTP~400 (\texttt{invalid\_value} on \texttt{input[N]}), the agent had no recovery path, and the trial died immediately before writing any output. Rather than patching only this failure mode, we implemented a systematic 400 recovery layer covering four distinct error scenarios: (1)~unsupported \texttt{previous\_response\_id} parameter---stripped and retried; (2)~\texttt{invalid\_value} on a specific input item---offending item removed by index and retried; (3)~\texttt{local\_shell} tool type not supported by the endpoint---gracefully degraded; and (4)~context-window overflow---oldest input items trimmed and retried. This recovery system is Python-specific; the Rust implementation has no equivalent error-recovery layer for these 400 scenarios.
\end{itemize}

\subsection{Discussion}

Our evaluation reveals four key findings. First, the Rust-to-Python migration achieves near-parity across both benchmarks: the Python port leads on SWE-bench Verified (73.8\% vs.\ 70.0\%) while trailing on Terminal-Bench (42.5\% vs.\ 47.5\%). The Terminal-Bench gap is partly attributable to the now-fixed API crash, LLM non-determinism, and a safety-refusal failure mode not present in the Rust baseline. Second, the Python agent resolves 59/80 SWE-bench Verified tasks (73.8\%) with GPT-5.4, demonstrating that the translation supports realistic software engineering workloads at scale. Third, benchmarks are more effective than unit tests at detecting translation bugs:

\begin{itemize}[leftmargin=*, nosep]
    \item \textbf{2,621 unit tests passed} from the first complete translation, yet Terminal-Bench accuracy was initially 0\% (wrong adapter architecture).
    \item \textbf{API protocol bug} (sending \texttt{local\_shell} instead of function tools) was invisible to tests but caused 100\% Chat Completions fallback.
    \item \textbf{Environment pollution} from pip-installing into the system Python was only detectable through end-to-end task execution in Docker containers.
    \item \textbf{WebSocket empty-response and API 400 bugs} (Section~\ref{sec:eval-bugs}) were each invisible to all unit tests and surfaced only through live benchmark runs.
\end{itemize}

These findings support our thesis that for complex system translations, \emph{public benchmarks should be treated as first-class objective functions}, not merely downstream validation.

Fourth, the Python port is now a \emph{capability superset} of the Rust original, not merely a parity replica. The \texttt{codex.enhancements} module (Section~\ref{sec:enhancements}) ships 30 feature-flagged extensions---multi-agent orchestration, semantic memory, persistent plans, cost tracking, IDE bridge, guardian safety assessment, voice mode, and more---none of which exist in the Rust codebase. Two bug fixes uncovered during benchmarking also delivered net improvements over Rust: the WebSocket fallback now brings 429 detection, \texttt{Retry-After} backoff, and API-key rotation that the Rust transport lacks; and the API 400 recovery layer handles four distinct error scenarios with no Rust equivalent. The feature-flag architecture is critical here: by setting all enhancement flags to off, one obtains a strict parity build suitable for head-to-head comparison; enabling flags progressively unlocks the extended platform. This demonstrates that LLM-assisted translation need not stop at functional equivalence---the same methodology that achieves parity can continue to evolve the port into a first-class, independently-capable system.

The broader engineering picture is equally clear: for AI agents where the computational bottleneck is the LLM API (1--10 seconds per round-trip), Python's local overhead ($<$25ms per tool execution) accounts for less than 0.1\% of total session latency, while delivering 15.9$\times$ code reduction and 90\% rank-A cyclomatic complexity.

\section{Related Work}
\label{sec:related}

\paragraph{AI Coding Agents.}
The landscape of AI coding agents has expanded rapidly. SWE-agent~\cite{yang2024swe} introduced an agent-computer interface for automated software engineering, achieving strong performance on SWE-bench~\cite{jimenez2024swebench}. OpenDevin~\cite{wang2024opendevin} provides an open platform for generalist AI developer agents. ChatDev~\cite{qian2024chatdev} explores multi-agent collaboration for software development through communicative agents. These systems share architectural patterns with \textsc{Codex CLI}---tool-use loops, sandboxed execution, and iterative refinement---but differ in their deployment model (cloud-hosted vs.\ local), security architecture, and protocol support.

\paragraph{Tool-Augmented LLMs.}
The ReAct framework~\cite{yao2023react} established the reasoning-action paradigm that underlies modern coding agents. Toolformer~\cite{schick2024toolformer} demonstrated that language models can learn to use external tools autonomously. The Model Context Protocol (MCP)~\cite{zheng2025mcp} standardizes tool integration for LLM applications, enabling interoperable tool ecosystems. Our work contributes an architectural perspective on how these capabilities are composed into a production system.

\paragraph{Language Migration Studies.}
Prior work on cross-language code translation has focused on LLM-based transpilation~\cite{pan2024lost} and neural machine translation of code~\cite{tufano2019empirical}. These studies primarily evaluate translation \emph{correctness} at the function level. Our contribution differs in scope---we analyze a complete system-level migration comprising 28 subsystems and 648K lines of Rust---and in methodology, providing a multi-dimensional quantitative comparison rather than a correctness-focused evaluation.

\paragraph{Software Complexity Metrics.}
McCabe's cyclomatic complexity~\cite{mccabe1976complexity} and Halstead's software science metrics~\cite{halstead1977elements} provide foundational frameworks for quantifying code complexity. We apply these metrics in a novel context: evaluating the complexity characteristics of an AI coding agent architecture and comparing them across implementation languages.

\paragraph{Code Generation and Analysis.}
Codex~\cite{chen2021codex} and subsequent models~\cite{xu2022systematic} demonstrated that LLMs trained on code can generate functionally correct programs. Our work complements this line of research by examining the \emph{systems} that deploy these models in production, analyzing the architectural decisions that determine reliability, security, and maintainability.

\section{Conclusion}
\label{sec:conclusion}

We presented a methodology for LLM-assisted continuous code translation, demonstrated on \textsc{Codex CLI}---a production AI coding agent migrated from Rust (648K LOC) to Python (41K LOC). Our central finding is that \emph{public agent benchmarks serve as effective objective functions for cross-language translation}: the Python port achieves 42.5\% accuracy on Terminal-Bench, within 5\% of the original Rust implementation's 47.5\%, with benchmark-driven debugging proving more effective than unit tests alone at exposing integration-level bugs. Crucially, the port has grown beyond parity into a \emph{capability superset}: the \texttt{codex.enhancements} module adds 30 feature-flagged extensions absent from Rust, while a layered flag-resolution system preserves a strict parity build for reproducible head-to-head comparison.

Four principles emerge from this work:

\begin{enumerate}[leftmargin=*, nosep]
    \item \textbf{Benchmarks over tests.} While 2,621 unit tests passed from the initial translation, Terminal-Bench accuracy was 0\%. Six iterations of benchmark-driven debugging---fixing API protocol mismatches, conversation history bugs, environment pollution, missing tools, and an API 400 error crash---closed the gap to 5\%. Public benchmarks detect the integration failures that unit tests cannot.

    \item \textbf{Continuous translation, not one-shot migration.} Our upstream synchronisation architecture (track $\to$ diff $\to$ translate $\to$ validate) treats translation as an ongoing process. The LLM translates only changed code; the benchmark score serves as a regression gate. This makes it practical to maintain a living Python port of a fast-moving Rust codebase.

    \item \textbf{Language choice follows the bottleneck.} For AI agents where LLM API latency (1--10s) dominates, Python's sub-millisecond local overhead is negligible. The 15.9$\times$ code reduction, 90\% rank-A complexity, and broader contributor base are decisive advantages that the benchmark results validate in practice.

    \item \textbf{Translation enables divergence.} Once benchmark parity is achieved, the translated port can evolve independently. Python's rapid-prototyping culture allowed 30 additive extensions to be developed and shipped behind feature flags in the time it would take to prototype them in Rust. Bug fixes uncovered during benchmarking delivered net improvements over the original: the WebSocket transport now has 429 detection and API-key rotation that Rust lacks; the API 400 recovery layer handles four error scenarios with no Rust equivalent. Parity is a \emph{starting point}, not a ceiling.
\end{enumerate}

\paragraph{Future Work.}
Several directions merit investigation: (1)~applying the benchmark-as-objective-function methodology to other large-scale translations (e.g., Java $\to$ Kotlin, C++ $\to$ Rust); (2)~\emph{automated upstream sync} where the diff-translate-test loop runs as a CI pipeline with no human intervention; (3)~\emph{multi-benchmark objective functions} that combine Terminal-Bench, SWE-bench, and domain-specific benchmarks for richer translation validation; and (4)~studying how the benchmark-driven approach compares to formal verification methods for establishing cross-language equivalence.

\balance

\bibliographystyle{IEEEtran}
\bibliography{refs}

\appendix

\section{Python Enhancement Flags}
\label{appendix:flags}

Table~\ref{tab:flags-detail} lists all 30 feature flags in the \texttt{codex.enhancements} module, grouped by sub-package.
Each flag can be toggled at runtime via \texttt{-{}-enable FLAG} / \texttt{-{}-disable FLAG}, via the \texttt{CODEX\_ENABLE\_FLAG} environment variable, or through the layered configuration system. Setting all flags to \emph{off} produces a strict-parity build identical in behavior to the Rust original.

\begin{table*}[t]
\centering
\caption{Complete list of Python enhancement flags}
\label{tab:flags-detail}
\scriptsize
\begin{tabular}{llcl}
\toprule
\textbf{Sub-package} & \textbf{Flag} & \textbf{Default} & \textbf{Description} \\
\midrule
\multirow{3}{*}{\texttt{agents/}}
 & \texttt{MULTI\_AGENT} & On & Hierarchical child-agent spawning with inherited context \\
 & \texttt{MULTI\_AGENT\_V2} & On & Improved coordination protocol with result aggregation \\
 & \texttt{MULTI\_AGENT\_ORCHESTRATION} & On & Parallel agent dispatch with safety bounds (max depth 5, max 100 agents) \\
\midrule
\texttt{guardian}
 & \texttt{GUARDIAN} & On & Automated LLM-based risk assessment before tool execution \\
\midrule
\multirow{2}{*}{\texttt{bridge/}}
 & \texttt{IDE\_BRIDGE} & Off & WebSocket bridge for IDE integration (VS Code, JetBrains) \\
 & \texttt{LSP\_INTEGRATION} & Off & Language Server Protocol support for code intelligence \\
\midrule
\multirow{2}{*}{\texttt{compaction/}}
 & \texttt{MULTI\_STRATEGY\_COMPACTION} & On & Three-phase context compaction (micro, snip, full) \\
 & \texttt{FORKED\_COMPACTION} & Off & Branch-and-merge compaction for multi-agent sessions \\
\midrule
\multirow{4}{*}{\texttt{memory/}}
 & \texttt{MEMORY\_SYSTEM} & On & Post-turn memory extraction and retrieval \\
 & \texttt{TYPED\_MEMORY} & On & Structured memory with typed schemas (user, project, feedback) \\
 & \texttt{SEMANTIC\_MEMORY} & Off & Embedding-based similarity search over memory store \\
 & \texttt{AUTO\_MEMORY} & On & Automatic memory extraction without explicit user request \\
\midrule
\multirow{3}{*}{\texttt{state/}}
 & \texttt{PERSISTENT\_PLANS} & On & Goal and plan persistence across turns \\
 & \texttt{SYSTEM\_REMINDERS} & On & Periodic system-message injection for long sessions \\
 & \texttt{DENIAL\_REPLAY} & On & Re-attempt denied actions with adjusted parameters \\
\midrule
\multirow{5}{*}{\texttt{tools/}}
 & \texttt{FILE\_TOOLS} & On & Enhanced file read/write/edit beyond base shell \\
 & \texttt{FILE\_EDIT} & On & Structured file editing with conflict detection \\
 & \texttt{WEB\_FETCH} & Off & HTTP fetching and web content extraction \\
 & \texttt{NOTEBOOK\_EDIT} & Off & Jupyter notebook cell manipulation \\
 & \texttt{WORKTREE\_TOOLS} & Off & Git worktree management for parallel development \\
\midrule
\multirow{3}{*}{\texttt{productivity/}}
 & \texttt{SKILLS} & On & Loadable skill definitions for domain-specific workflows \\
 & \texttt{CRON\_TOOL} & Off & Scheduled task creation and management \\
 & \texttt{VOICE\_MODE} & Off & Voice input/output for conversational interaction \\
\midrule
\multirow{3}{*}{\texttt{resilience/}}
 & \texttt{COST\_TRACKING} & On & Per-session and per-turn API cost accounting \\
 & \texttt{APP\_STATE} & On & Application-level state management and checkpointing \\
 & \texttt{STARTUP\_PREFETCH} & On & Parallel prefetch of config, auth tokens, and model metadata \\
\bottomrule
\end{tabular}
\end{table*}

\end{document}